# The informational model – possible tests

S.V. Shevchenko[1]  and V.V. Tokarevsky[2]

[1]*Institute of Physics of NAS of Ukraine, Kiev, Ukraine*
[2]*State corporation "Radon",   Kiev, Ukraine*

**Abstract**   In a number of our arXiv papers (more systematically the informational conception is presented in the paper "The Information as Absolute", 2010) it was rigorously shown that Matter in our Universe – as well as   Universe as a whole - are some informational systems (structures), which exist as uninterruptedly transforming [practically] infinitesimal sub-sets in the absolutely infinite and fundamental set "Information". Such a conception    allows suggesting a reasonable physical model that is based on the conjecture that Matter is some analogue of computer (more correct – of a [huge] number of mutually comparatively independent automata). The conjecture, in turn, allows introducing in the model the basic logical elements that constitute the material structures and support the informational exchange - i.e. the forces - between the structures.  The model  yet now makes  more clear a number of basic problems in special relativity, quantum mechanics, and, rather probably, in [now – in Newtonian] gravity. In this paper some possible experiments for the model testing are considered

## 1. Introduction

In [1 - 3] it was rigorously shown that Matter in our Universe – and   Universe as a whole - are some informational systems (structures), which exist as uninterruptedly transforming [practically] infinitesimal sub-sets in absolutely infinite and fundamental set "Information". This informational conception allows to propose the physical model (more see [4]), which, when basing *practically only on Uncertainty principle*, adequately depicts the motion and interactions of particles in spacetime.  In the model a [subatomic] particle is some closed – loop algorithm that runs on a "hardware", which, in turn, consists of a closed chain of elementary logical gates – fundamental logical elements (FLE). The FLE's sizes in both  - in the space and in the time - directions are equal to Planck length, $l_P$, $l_P = (\frac{\hbar G}{c^3})^{1/2}$ ($\hbar$ is reduced Planck constant - the elementary physical action, $G$ - gravitational constant, $c$- speed of light in the vacuum); minimal time of the FLE's "flip" is equal to Planck time, $\tau_P$, $\tau_P = \frac{l_P}{c}$. Below in this section we give some brief introducing in the informational model to understand the main text. If, nonetheless, some questions occur, then see [4].

### 1.1. Particles

Since particles' algorithms never stop (and the FLEs are uninterruptedly flipping), it becomes be rather reasonable to introduce the "informational currents" (IC) and fixed information variables:
- the time IC (t-IC):

$$j_t = \frac{1}{\hbar} \gamma m_0 c^2, \quad (1)$$

- the space IC (s-IC):

$$j_x = \frac{1}{\hbar} \gamma m_0 c^2 \beta^2, \quad (2)$$

- the fixed information:

$$\Delta I_M = \frac{\Delta M}{\hbar}. \quad (3)$$

($v$ is the speed of a particle, $\beta = v/c$, $\gamma = 1/(1-\beta^2)^{1/2}$ is the Lorentz – factor of the particle motion, $\Delta M$ is the angular momentum, $m_0$ is the particle's rest mass. The dimensionality of the time and the space currents is [bit/s], the dimensionality of fixed information is [bit]). Besides note that fixed information relates, quite naturally, also to the physical action, $S$.

The "material" length of a particle's algorithm [at rest] is equal to the particle's Compton length, $\lambda_C$, $\lambda_C = \frac{\hbar}{m_0 c}$.

So through a particle's circular logical chain an active "flipping point" runs uninterruptedly, having momentum, $p_P$, $p_P = \hbar/\lambda_C = m_0 c$, and angular momentum (for example – the photon's spin) $\hbar$.

If in spacetime a flipping point runs through a straight line (in a space or in the time direction), then some impact with momentum $p$ in this direction results in occurrence of a particle – at the impact in the time direction that is "usual" material particle ("T-particle") having the mass $m_0 = p'/c; p' \leq p$; the impact in a space direction results in the occurrence of "X-particle", e.g. – of a photon having the energy $E = pc$.

Any [of known now] particle's Compton length is much larger the Planck length, what allows "to write" on this length a code that defines the particle's parameters, but all (any particle's) codes contain "universally significant" FLEs - "us-FLEs", that flip in the end of the algorithm, i.e. in the end of particle's Compton length. And just these FLEs determine the location of the particle in spacetime, besides it is rather probable (see below) that these FLEs responsible also for the gravity interactions between particles (and, of course, – between bodies). Note also the important feature of the FLEs – to flip without dissipation of an energy they should be logically reversible (some analogies of "Toffoli gates"). Then it is reasonable to suggest that particles and corresponding antiparticles are some logical structures that have opposite sequences of commands in their algorithms.

**1.2. Spacetime**

Introducing of the Space and the Time notions in the model are quite natural – they are some conditions (rules) that allows (and define how to single out) to single out specific informational patterns / structures - e.g., particles - in the main structure (i.e., - in Matter) at that taking into account both - fixed and dynamical – characteristics of the structures. As some rules Space and Time are "absolute" and exist "forever", since they exist also ("virtually") before a Beginning and after an End of some specific informational structure, for example – of our Universe. After "materialization" at Beginning, Space and Time remain be absolute, revealing themselves as "the time" and " the space" variables, when any element of the structure – a particle, a molecule, a star, etc. – have its own (*individual, proper*) space and time parameters in *absolute spacetime*. A particle always moves in spacetime - in the time and / or in the space directions. Both directions are in many respects equivalent but aren't totally identical. First of all – logically any step in the space is simultaneously the step in the time. That is the cause, rather probably, of a number of differences in the particles characteristics.

Above one difference was pointed out – impacts on straight line flipping FLE in different directions result in occurrence of different (T- and X-) particles. Besides – when the time direction is unique, a particle can move in space in 3 different directions; T-particle can move along some open lines in the space and the time directions, when X-particle can move only in a space direction, with the flipping point moving back and forth in time direction; and – if a photon has "correct" spin in the space, the majority of T- particles have non –integer "space" spins – though it is rather probable that in the time direction a T- particle's angular momentum is "correct", i.e. is equal to $\hbar$.

Both - the "time dilation" for T-particles and the (generally speaking – independent on the time dilation) "space dilation" for X-particles [that occur at particles' motion] have the same cause –   that the FLE's flip rate cannot be lesser then inverse Planck time. In Nature that reveals as the fact that *measured* speed of light is constant in inertial reference frames moving with different speeds. This fact was introduced in special relativity theory (further - SRT) as fundamental "second postulate", resulting in a number of the SRT's inconsistencies, first of all - in the inference that   "at a motion of a frame the space transforms into the time and vice versa".  In the informational model   that isn't so. Since any particle (or a system of interacting particles) has its own specific time and space parameters in the absolute spacetime of Matter, a motion of a particle affects only upon the parameters of this particle (system of particles) and nothing does with the external Matter, including – nothing does with spacetime.

### 1.3. Forces in the informational model

In the informational model seems as quite plausible the conjecture   that at an interaction of a force's mediator with a particle some t-IC step in this particle becomes "be spent" by interaction, resulting in the particle's t-IC's decrease (if potential energy, $U<0$) and in corresponding mass defect; or "be added" resulting in t-IC increase if $U>0$. Correspondingly at the interaction the mediator transmits to the impacted particle a momentum, $\vec{p}_0$.

## 2. The experiments

### 2.1. Gravity model testing

It is possible to put forward, [1] rather reasonable conjecture - since   the gravity force is universal (regardless to the kind of particles) - that the gravitational potential energy of a system of some bodies is proportional to the accidental coincidence rate of some equivalent of the t-ICs of the particles of these bodies. Such coincidences always exist since the t-FLE's (the particle's FLE's) flip-time is not equal zero. Secondly suppose  that  in gravity interaction only us-FLEs, i. e. the FLEs that are used for localization of particle in space, "take part".

   Basing only on approach of section 1 and the conjectures above, the equation for potential gravitational energy can be obtained as follows.

    As that was assumed above, the t- and s-FLE's (space FLEs or "aether" FLEs) sizes are equal to Planck's length, $l_P$. Besides assume that:

(i) - at every t-IC step of a particle in   space a "rim" of s-FLE's flips starts to expand with radial speed that is equal to the speed of light, $c$, so the rim's area is equal $2\pi r l_P$  ($2\pi c t l_P$);

(ii) - the time of the t-FLE's flip, $\tau_t$, and of the interaction of the s-FLEs and t-FLEs, $\tau_r$, are the same $\tau_t = \tau_r \equiv \tau = l_P/c$, i.e. – and  are equal to Planck time;

(iii) – for the information decrement in the t-IC be equal [in this case, i.e. $U<0$] to "$-\hbar$" is necessary two such interactions.

So the [average] accidental coincidence rate in the particle 2 when radiates the informational current of the particle 1, $N_{cc21}$, is equal:

$$N_{cc21} \approx \frac{j_{t1} \cdot 2\pi r l_P}{4\pi r^2} \cdot P \cdot j_{t2} \cdot 2\tau = \frac{m_1 c^2 \cdot 2\pi r l_P}{4\pi r^2 \hbar} P \frac{m_2 c^2}{\hbar} 2\tau, \qquad (4)$$

where $P$ – is the probability of particle's 2 us-FLEs interaction if a rim of the particle's 1 t-IC's s-FLE flips passes through this us-FLE.

Since the system is symmetrical, the coincidence rates for both bodies are equal and the potential (binding) gravitational energy is equal (for $P=1$):

$$E_{gI} \approx -\hbar \cdot N_{cc21} = -\frac{c^3}{\hbar r} l_P^2 m_1 m_2 = -\frac{l_P^5}{\hbar r \tau^3} m_1 m_2 = -G \frac{m_1 m_2}{r} \qquad (5)$$

From Eq.(4) evidently follows the equation for the gravity force ($\vec{p}_0 = -\hbar \vec{r}/r^2$) in the statics:

$$\vec{f}_g = \frac{d\vec{p}}{dt} = N_{cc21} \vec{p}_0 = -\frac{G m_1 m_2}{r^3} \vec{r}. \qquad (6)$$

From Eq. (4) follows also that the gravitational (coincidence rate) current, $j_g$, in any particle is random in the time and its average value is

$$j_g = G \frac{m_1 m_2}{\hbar r} \qquad (7)$$

Correspondingly so does the gravitational force that impacts (as a sequence of elementary momenta, $p_0 = \hbar/r$) on this particle and some gravity force's randomness should occur at interaction of small masses. The detection of this randomness (or some equivalent physical value) would be rather weighty evidence that suggested informational model is true.

**2.1.1. Deviation of neutrons in test mass gravity**

The schematic setup (not to scale) of possible experiment for measuring of the deviation of ultracold neutrons in a test mass gravity is shown in Fig. 1. Collimated neutrons from the UCN source, S, run trough the slit in a screen. On the way "the slit – position sensitive detector (PSD)" the parallel beam is distorting – at the diffraction (on the angle $\vartheta$ in the figure) and at the gravity force impacts (on the angle $\varphi$) that are created by the test mass TM. The neutrons hits are detected in the PSD depending on the distance from the beam axis to the hit, $z$. It is evident that the displacement and displaced fraction of initial UCN beam under the test mass gravity rises when the speed $V$ decreases, but that give rise also to increasing of the diffraction angle. On the other hand one can reduce the diffraction background by increasing slit's width, but at that appears a probability of the background increasing because of a contribution of the next diffraction maxima. Another difficulty in possible experiment is the growth of experimental efforts (and the cost) to achieve the neutron's cooling behind 1m/s (~50μK).

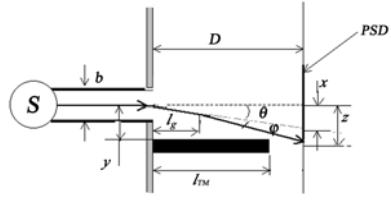

Fig. 1. Schematical setup of the experiment for measuring the UCN deviations in the test mass gravity. The notations – see the text.

So the calculations were carried out for the speed $V=1$ m/s, tungsten test mass having thickness 10 mm and length, $l_{TM}$, 0.9m.

It can be easily shown that for small distances "a spherical test mass – a particle", $l$, the momenta, $p$, impacting on the particle towards the TM's centre are in the range $p \in (\frac{1}{l}, \frac{1}{2R})$ for the sphere's radius $R$ (see Fig.2).

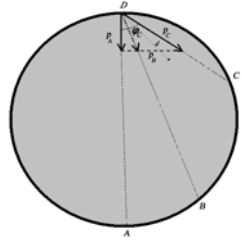

Fig. 2. Momenta at the gravity interaction of fragments of a spherical mass and a particle which is placed near point D.

Indeed, minimal – and equal – momenta' values correspond to the test mass's surface fragments:

$$p_S = \frac{\hbar Cos(\varphi_G)}{d} = \frac{\hbar}{2R}, \qquad (8)$$

where $d$ is, e.g., the distance [CD] on the Fig.2; momenta for fragments inside sphere are lager, $p > p_S$.

At every impact the neutron starts motion with the speed $v = p/m_n$, where $m_n$ is the neutron's mass; so the neutron's trajectory becomes be deflected on the angle $\varphi, \varphi \approx p/P$, where $P = m_n V$, $V$ is the neutron's initial speed, $v \perp V$.

Besides, it is well known from textbooks that if neutrons pass through a slit having width $b$, the parallel neutron beam becomes be distributed in a section (plane) as

$$Q \propto \frac{Sin^2(u)}{u^2} \qquad (9)$$

where $u = \frac{\pi b}{\lambda} Sin\vartheta$, $\lambda = h/P$ (~400nm for $V$=1m·s$^{-1}$) is the neutron's de Broglie wave length, $Sin\vartheta \approx x/D$ if $D\gg x$, where, in turn, $x$ is the displacement of neutron's hit point from the beam's axis; $D$ is the distance from the slit to the plane (see Fig. 1).

Some examples of the neutrons' [diffraction] distributions are shown on Fig. 3 for the neutrons' speed $V = 1 m \cdot s^{-1}$, and a number of the slit widths, $b$=0.4, 0.8, and 1.6 mm.

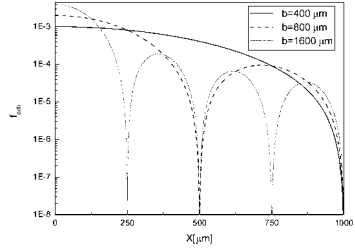

Fig. 3. Probability densities for neutrons' hits in the PSD plane at the diffraction for a set of slit's widths, $b$; $b$=0.4, 0.8, and 1.6 mm. The distributions are normalized on the neutrons' flux ½ °n.

On Fig. 4 the distributions of $x$ values under the TM gravity force impacts for 10 distances "TM – the UCN beam axis", $y$; $y$=50, 100,…500 μm are shown. Since the distributions don't differ essentially, in the calculation further the distribution for $y$=200 μm was used.

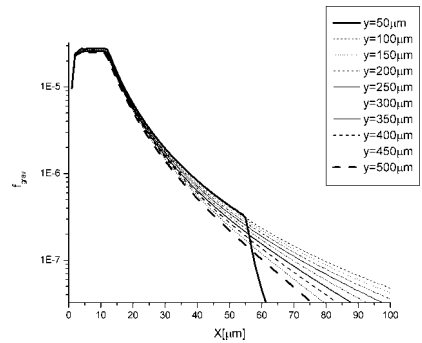

Fig. 4. Probability densities for neutrons' hits in the PSD plane under the test mass gravity impacts for a set of the distances "the TM – the UCN beam axis", $y$, $y$=50, 100,…500 μm and the TM's length 0.9m. A knee on the "50μm" line is because of the TM's shadow. The distributions are normalized on the neutrons' flux 1.0 °n.

If both factors (the diffraction and the TM gravity) act on neutrons, the diffraction pattern becomes be "smeared" towards the TM. The difference of resulting distributions (when both factors and only the diffraction affect upon neutrons) are shown in Fig. 5 for the slit's width values $b$ given above. For these $b$ values the "signal - background" ratios, $SB$ ($S$ and $B$ are integrals of the distributions in the intervals from 0.9 to 0.1 of the first maximum of a distribution), are: $SB \equiv S:B =$ (5.3·10$^{-6}$ : 4.1·10$^{-1}$), (1.0·10$^{-5}$ : 3.6·10$^{-1}$) and (1.7·10$^{-5}$ : 2.6·10$^{-1}$). As it is expected, the best $SB$ value occurs for the maximal slit's width, $b$=1.6 mm, when $SB$

~$6.5 \cdot 10^{-5}$. This value is rather small so for typical $3\sigma$ deviation from the diffraction background the neutron flux ($\approx 9B/S^2$) ~$10^{10}$ °n should be used.

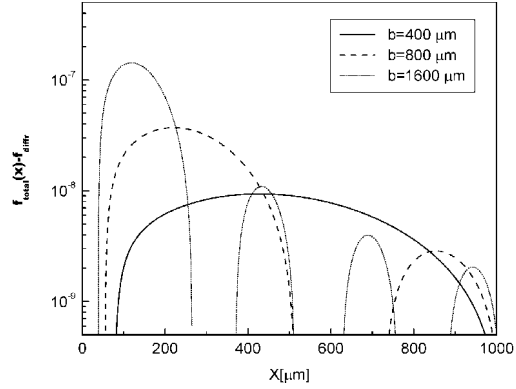

Fig. 5. The difference of resulting distributions (when both factors and only the diffraction affect upon neutrons), $f_{total}$ and $f_{diffr}$, for the slit's width values $b$ =0.4, 0.8, and 1.6 mm.

### 2.1.2. Monochromatic photon beam distortion

Utmost "lightweight" – and available - particles are photons, so in this case the randomness of the gravity force can be observed when Earth is used as a test mass. Corresponding experiments were considered earlier, [4, 6] including the caution that photons are principally relativistic particles and so the randomness' estimation with using the equations Eqns. (5,6) for gravity force (which are valid if both bodies are at rest) can be not totally correct; but for a convenience we present here the results.

So if $m_1$ and $m_2$ in Eq. (7) are Earth and a photon having frequency $\nu_0$, then the rate of gravity impacts on the photon, $n_\gamma$, is

$$n_\gamma = 2\pi G \frac{m_1 \nu_0}{R_E c^2} \qquad (10)$$

As that was shown above, the minimal momentum impacting on a particle is $p_b = \hbar/2R$ if the body 1 is a sphere with radius $R$ and the distance "body 1 – the particle", $l$, is small, $l \ll R$. In zero approximation, for Earth's radius, $R_E \approx 6.35 \cdot 10^6$ m and mass, $M_E \approx 6 \cdot 10^{24}$ kg, near Earth surface corresponding rate is: $n_\gamma \approx 4.4 \cdot 10^{-9} \nu_0 s^{-1}$.

Since a photon's initial momentum is $p_p = h\nu_0/c$, at every gravitational interaction the photon's frequency becomes be shifted [on $\Delta\nu = c/2\pi r$] at least on the value $\Delta\nu = c/4\pi R_E \approx 3.76$ Hz if the photon beam is normal to Earth surface.

Thus on the way *l* average number of "photon - Earth gravity" interactions, *N*, is $N = n_\gamma t = n_\gamma \dfrac{l}{c}$ when number of the gravitational interactions of a photon in this distance, *k*, is a random and is distributed under Poisson law:

$$p(k) = \dfrac{N^k e^{-N}}{k!} \qquad (11)$$

Besides, since the paths, where an interaction happens, and transmitted momentum's values [since Earth is rather large body] are random, the part of interacted photons in initial beam becomes be incoherent.

In contrast to this informational model, General Relativity theory , [9], operates with continuum spacetime and so, according to the GR, in this case frequencies of *all* photons must be coherently "shifted" (see section 3) on the equal value, $\Delta \nu_{GR}$:

$$\Delta \nu_{GR} \approx \nu_0 \left(\dfrac{\varphi_0 - \varphi_l}{c^2}\right) \approx \nu_0 \dfrac{GM_E}{(R_E c)^2} l \approx 10^{-16} \nu_0 l \qquad (12)$$

(Note, however, that the consideration above is incorrect in some sense – if the GR is true then a photon doesn't interact with the gravity (and so doesn't change its frequency): observed (e.g. - [10], [11]) red/blue "shifts" are results of different time dilation values for radiating and "detecting" atoms, if they are in different gravitational potentials. But, on the other hand, experiments show that photons should interact with the gravity – if a portion of photons is placed in a box having reflecting inner surfaces, then the gravitational mass of this box is larger then the mass of the empty one. The suggested experiment allows clearing up this problem to some extent.)

In the table 1 some estimations of incoherent - and having different frequencies – photon fractions in the initial beam are given for two frequency limits - $\Delta \nu > 3.7$ Hz and $\Delta \nu > 200$ Hz. In the last case corresponding radius is equal $\sim 1.2 \cdot 10^5$ m, the mass (for average Earth density 5 g·cm$^{-2}$) $\sim 3.6 \cdot 10^{19}$ kg, $n_\gamma \sim 1.4 \cdot 10^{-12} \nu_0 \text{s}^{-1}$.

Table 1. The incoherent fractions in the initial vertical (to Earth surface) monochromatic photon beam which appear after passing the way, l, and which have the dispersed frequency that differ from initial value more then on Δν in Earth gravity field. The data for coherent GR frequency shift are shown also.

| $l$ [m] | H-maser (22 cm) $\nu_0$[Hz]= 1.4 $10^9$ | | | H2O maser $\nu_0$[Hz]= 2.2 $10^{10}$ | | | 698 nm [7] $\nu_0$[Hz]= 4.3 $10^{14}$ | | |
|---|---|---|---|---|---|---|---|---|---|
| | Δν>3.7 [Hz] | >200 [Hz] | Δν$_{GR}$ [Hz] | Δν>3.7 [Hz] | >200 [Hz] | Δν$_{GR}$ [Hz] | Δν>3.7 [Hz] | >200 [Hz] | Δν$_{GR}$ [Hz] |
| 1 | 2 $10^{-8}$ | 7· $10^{-12}$ | 1.4·$10^{-7}$ | 3· $10^{-7}$ | 1· $10^{-10}$ | 2.2·$10^{-6}$ | 6·$10^{-3}$ | 2·$10^{-6}$ | 0.043 |
| 10 | 2· $10^{-7}$ | 7· $10^{-11}$ | 1.4·$10^{-6}$ | 3· $10^{-6}$ | 1· $10^{-9}$ | 2.2·$10^{-5}$ | 0.06 | 2·$10^{-5}$ | 0.43 |
| 100 | 2 $10^{-6}$ | 7· $10^{-10}$ | 1.4·$10^{-5}$ | 3· $10^{-5}$ | 1· $10^{-8}$ | 2.2·$10^{-4}$ | 0.6 | 2·$10^{-4}$ | 4.3 |
| 1000 | 2· $10^{-5}$ | 7· $10^{-8}$ | 1.4·$10^{-4}$ | 3· $10^{-4}$ | 1· $10^{-7}$ | 2.2·$10^{-3}$ | 100% | 2·$10^{-3}$ | 43 |
| $10^4$ | 2· $10^{-4}$ | 7·$10^{-7}$ | 1.4·$10^{-3}$ | 3· $10^{-3}$ | 1· $10^{-6}$ | 2.2·$10^{-2}$ | | ~2% | 430 |
| $10^7$ | ~ 20% | ~ $10^{-5}$ | ~0.9 (*) | ~0.5 | ~2 $10^{-4}$ | ~14 | | | ~5·$10^5$ |

(*) – there was the attempt to measure this GR shift 0.9Hz (R.F.C. Vessot et. al., Phys. Rev. Lett., V45, No 26, 980, P 2081; but they used a signal filtration with "necessary" filter's band and 100 s averaging time.

## 2.2. SRT testing

In the informational model Lorentz transformations can be obtained quite naturally, [4]:

the first equation

$$x = vt + x'(1-\beta^2)^{1/2}, \qquad (13)$$

and the second one:

$$t' = (1-\beta^2)^{1/2} t - \frac{vx'}{c^2}, \qquad (14)$$

but with essential difference from standard SRT – these equation aren't valid in whole spacetime but are true *inside "a moving wagon"* (see Fig. 6) *only*: $x' \in (0, L)$, $x \in (x_0, x_1)$; $x_0 = Vt$, and $t' \in (t'_0, t'_1)$; $t'_0 = t(1-\beta^2)^{1/2}$.

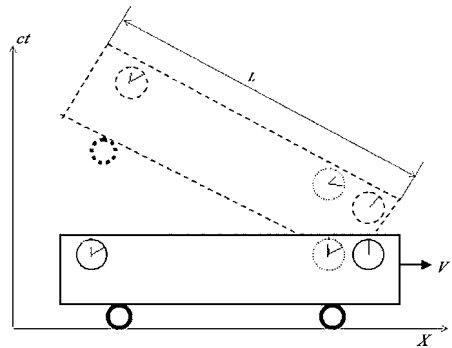

Fig. 6. A wagon having at rest the length $L$ moves with speed $V$ along X-axis.

The $t$- decrement for the wagon's matter along the wagon's length (the maximum is $-\frac{VL}{c^2}$), appears at the acceleration of the wagon up to the speed $V$ and further remains be constant for any fragment of the wagon at the uniform motion. So if one synchronizes a couple of clocks in the ends of the wagon before the acceleration, then he always can measure the wagon's speed relative to the starting reference frame (RF), when that is impossible in standard SRT. From Eq. (14) follows, that if a wagon has the length 100m and if the wagon is accelerated up to speed 100m/s (360km/h), then the decrement will be ~ $10^{-13}$s – the value that, rather probably, now can be measured – after somebody has moved, e.g., the clock in the left end to the right end of the wagon (or has co-moved both clocks together in any place of the wagon) with speed $v<<V$ and compare the clocks' times.

## 3. Discussions and conclusion

Above two experiments aimed at the detection of the gravity force randomness are depicted. They have some advantages and disadvantages. The experiment with the UCN is more complicated comparing with the photon beam distortion measurement, but, since neutrons are practically at rest, the gravity model in this case can be applied directly. The experiment with photons seems as simpler – the sub-hertz linewidth lasers (and the masers) aren't now some exotics (see, e.g., [7, 8]), but, as that was pointed out earlier, photons are principally relativistic particles and so the beam distortion estimations in Section 2.1.2 can be, rather probably, only some zero approximations of the effect. Nonetheless some effect must reveal itself if the suggested gravity model is true at all.

Both experiments are rather difficult when at first sight given in Section 2.1 gravity model seems as rather speculative. However, since:
 (i) – the model quite naturally follows from the informational conception, when existence [of the Set "Information" and of Matter as of an informational structure in this Set], truth and self – consistence of this conception are rigorously proven;
(ii) - Eq. (5) by any means doesn't follow from both – from [experimental] Newton's gravity law and from Planck's approach at obtaining his "natural units", thus it seems as non - accidental and so there is non- zero probability that Eq. (5) is true;
(iii) - the informational model allows to make essentially more clear a number of basic physical problems,
- the proposed in Section 2 experiments are worthwhile to be done.

As well as these experiments can give an evidence for another possibly interesting sequence from the model. As it is known from a number of experiments, the gravity force acts on matter and antimatter particles identically, from what follows that the us-FLEs in these both cases have, rather possibly, identical – and so totally symmetrical – logical structures. In [2] it was conjectured that at Beginning of our Universe the first Matter's particles were utmost simple, i.e. that were Planck mass particles (PMP). These particles have very simple logical structure – their FLE chain contains only us-FLEs and so there isn't place to write any information besides that defines the localization of a PMP in spacetime and its gravity interactions with other particles. Just after Beginning, when the density of PMPs was very large, a part (~25%) of the PMPs interacted with the creation of other Matter particles – baryons, leptons, photons, etc.; when the rest (~75%) exist till now as "dark matter" particles.

This conjecture can explain the nature of the dark matter and its evident properties – the participation only in gravity interactions in large space scales and the absence in reality of interactions in the [WIMP] detectors' scales because of extremely low PMP concentration in the space and comparatively low probability of gravity interactions yet at not too high temperature.

But from the symmetry of the us-FLEs follows also another sequence. Indeed, since the us-FLEs are totally symmetrical, the notion "reversibility" looses a sense. As well as the difference does between particles and antiparticles (as "left- rotated" and "right- rotated" particles) – all (PMP or some other totally symmetrical initial particles) matter at Beginning was "one sort" – say - "right- rotated" matter. It seems quite naturally further to conjecture that decays of the "right- rotated" PMPs resulted further in occurrence of [practically] only "usual" matter in our Universe.

The fact of the creations in particle-particle reactions of the pairs "particle – antiparticle", is, rather possibly, only some "by-product" of the reversibility – and of non- symmetry – of other particles' FLE codes. Though note that this "by-product" conserves the matter's quantity in Universe in certain sense.

The standard SRT version contains some self-contradictions, e.g., - well known "the twin paradox". Existing "resolution" of the paradox as that it arises because of the twin-traveller's reference frame is non- inertial - when homebody's one is inertial - is evidently unsatisfactory. Besides - it seems as not too plausible to think that every moving particle (which has, of course, its own inertial frame), e.g., - in an accelerator - "transforms" whole spacetime in Universe. The standard SRT evidently is non – consistent with the Big Bang hypothesis – again it seems as not too plausible to think that, e.g., every electron in LEP enlarges the energy of Matter in Universe (and so – Big Bang energy; though Big Bang has happened already rather long time ago) in ~$2 \cdot 10^5$ times.

The informational model states that any impact on a particle (a "rigid" system of particles) changes the parameters of this particle (system) only, including – slows down / increases the particle's algorithm operation rate relating to the absolute time (i.e. slows down / increases the "time dilation" for T - particles); increases / decreases the particle's mass (energy), etc. When in all other respects the informational model is analogues to the standard SRT – since for the both the same Lorentz transformations are valid - the model hasn't the SRT flaws pointed out above. The experiment in section 2.2 of this paper would allow convincible testing of the model (and of the SRT, of course), at that it is possible yet now by using existent techniques – the speed of some trains is now near 400 km/h, when the time can be measured with accuracy ~$10^{-16}$ (see, e.g., [12]).

Though in this case the result can be zero and the clocks will show equal times after co-moving, since the "wagon" as a whole, i.e. including the clocks, constitutes *a rigid system* where all parts mutually interact and at the co-moving process the time parameters of the clocks will change depending on clocks' position ($X$ coordinate). So more correct version of the test is when the co-moving proceeds outside the wagon, for example – in a wagon that moves trough parallel rail with a (practically) same speed and so is in the same reference frame as the first one. Such a procedure seems possibly too difficult when two high-speed trains on Earth are used, but seems simpler, e.g., for a pair of space satellites.

The suggested tests relate, in fact, to well-known clocks synchronization problem in the SRT, particularly the section 2.2 test version above relates to the problem of equivalence of "Einstein" (when clocks are set in accordance with the convention that speed of light is constant in any inertial RF) and "slow clock transportation" methods of the synchronization. So if the clocks inside the first wagon show equal times after co-moving than a next confirmation of these standard enunciations of the SRT occurs (but doesn't contradict with the model). But if the clocks inside the "parallel" wagon don't show equal times after co-moving, that would be the confirmation of the informational model.

So the model testing rather probably can be reduced to the checking of equivalence of spatially separated clocks in "rigid" and "independent" systems when systems as a whole move with the same speed. One of possible experiments then can be done as follows.
Let there are two clocks, a pulsed light source (PLS) and a mirror (M) in a satellite, and, after clocks synchronization, one of the clocks and the mirror start moving along the satellite orbital motion with (relative) speed $v<<V$, where $V$ is orbital speed of the satellite. In this case the clocks pair, rather probable, doesn't constitute a "rigid" system, but clocks move in the same RF. After the distance, $L$, between clocks will become be large enough, the PLS emits light flashes which, after reflection from the mirror, return to the satellite. At that both clocks log out the times of the light's emitting, $t_1$, reflection, $t_2$, and returning, $t_3$. After a number of flashes the logs are compared.

If the SRT is true, then intervals ($t_{2i}$- $t_{1i}$) and ($t_{3i}$- $t_{2i}$), $i$=1,2,3…, must be equal; but if that isn't so, then the intervals will differ on the value $\approx 2\dfrac{VL_i}{c^2} \approx \beta(t_{3i} - t_{1i})$.

Note, that very probably the difference will depend on just the speed $V$, not on Earth speed in absolute space; it is very probable that under gravity force the system "Earth – satellite - clocks" constitute a rigid system. So to measure the "absolute speed", as it seems, is necessary to send the satellite with clocks somewhere in space where the gravity would be weak enough…